**This manuscript is a pre-print only version, if you find any error, please, kindly send me an email to [agomez@icmab.es](mailto:agomez@icmab.es). Thank you very much for your comprehension.**

# Resetting Piezoresponse Force Microscopy: Towards a real quantitative technique

A. Gomez[1]*, H.T.T Nong[2], S. Mercone[2], T. Puig[1], X. Obradors[1]

[1]Instituto de Ciencia de Materiales de Barcelona (ICMAB-CSIC), Campus UAB, 08193, Bellaterra, España
[2]Laboratoire de Sciences des Procédés et des Matériaux (LSPM-CNRS UPR-3407), Université Paris 13, Sorbonne Paris Cité, 99, Av. J. Clément, 93430 Villetaneuse, France

*Corresponding author: agomez@icmab.es

ABSTRACT

A nanometric needle sensor mounted in an Atomic Force Microscopy allows systematic picometer-range distance measurements. This force sensing device is used in Piezoresponse Force Microscopy (PFM) as a distance sensor, by employing the cantilever spring constant as the conversion factor opening a pathway to explore the piezoelectric effect at the nanoscale. The force-distance equivalence is achieved if the force sensor does not disturb the system to study, solely. In this manuscript we report a systematic study in which different Lead Zirconate Titanate (PZT) materials, having different $d_{33}$ values, are measured following the standard theory available for PFM. Both in resonance and out of resonance measurements demonstrate that PFM cannot be considered quantitative. After performing the measurements, we propose a correction of the standard theory employed in PFM by considering the force exerted by the material as a variable. The $g_{33}$ parameter, inherent to piezoelectricity, governs the amount of force available from the system. A comparison of piezoelectric stiffness's for the case of a nanoscale site contact region, similar to the one it is found while performing PFM, is provided. Such stiffness is well below the cantilever stiffness, limiting and diminishing the material movement, as the piezoelectric material does not have enough stroke to induce the intended displacement. A correction factor, named Open Piezopotential Gauge, accounts for these effects, which is used to correct the measurements carried out in PZT materials towards a real quantitative PFM.

MAIN TEXT

Piezoelectricity and ferroelectricity are extensively investigated physical phenomena since its discovery in 19[th] century[1–3]. Applications of piezoelectricity account from a typical fire lighter to car ignition systems[4]. Thus, from specific applications to everyday ones, their use extends progressively[5–8]. Indeed, for the growing energy demand as well as for the storage of the information technology, piezoelectric small materials seems to be very attractive to power up portable piezoelectric nanogenerators[9], as well as new magnetoelectric smart high density memories[10]. This kind of development brings up the need for understanding at a nanoscale level the ferroelectric phenomena. This kind of development brings up the need for understanding at a nanoscale level the ferroelectric phenomena. Thanks to the technical advances in microscopy probes, the possibility to switch locally the polarization of a ferroelectric thin film for a possible ultrahigh density information device have been proposed early in this century[11]. As a matter of fact, reducing the volume of the device has consequences on the optimized macroscopic piezoelectric properties[12,13]. Thus to quantitatively characterize and/or control the good piezoelectric local properties of these nano-devices is of primary importance for their future applications[14–18].

Piezoresponse Force Microscopy (PFM) arises as an advanced characterization mode based on Atomic Force Microscopy (AFM) capable of locally characterizing piezoelectric and ferroelectric materials[19–21]. In such mode, a metallic tip is engaged into contact with a piezoelectric material surface, while at the same time, an AC bias is applied through the tip or the sample[22]. An electromechanical vibration is induced due to the inherent piezoelectricity, at the same frequency. The amplitude of vibration of the tip, out of the resonance, is[23–25]:

$$A_m = d_{33} V \qquad (1)$$

Where $A_m$ is the tip amplitude [$m$], $d_{33}$ is the piezoelectric constant of the material [$mV^{-1}$] and V the applied bias amplitude [$V$]. Working at the contact resonance enhances the tip vibration by the quality (Q) factor of the resonator, through the following equation[19,26]:

$$A_m^* = A_m Q \qquad (2)$$

Both expressions describes PFM as a quantitative method where the $d_{33}$ parameter can be estimated from the electromechanical behavior. Although many researchers active in the ferrolectric/piezoelectric community, adopted PFM as a characterization tool able to proof such electrical, a huge controversy on the results obtained is nowadays well established. This latter mainly concerns the reliable into quantitative analysis of the PFM results, as denoted during the annual largest conferences in Piezoelectrics (ISAF/ECAPD/PFM Conference 2016) by Alexei Gruverman. In this manuscript, we report measurements of the $d_{33}$ piezoelectric constant of several piezoelectric materials, all of them made of Lead Zirconate Titanate (PZT) supplied by Morgan Advanced Materials. In these compounds the variation of composition influences the $d_{33}$ value[27]. Each of the material measured has been previously characterized by macroscopic measurements. These latter are used as a comparison for the PFM local results. Here we prove that PFM cannot be considered, in its actual analysis form, a quantitative technique. According to our measurements, we introduce an opportune correction factor in Equation (1) that we will call, the Open Piezopotential Gauge, $ɣ_{OPG}$, to compile a reliable $d_{33}$ determination. This constant is related to the maximum force that the piezoelectric material can exert, governed by the piezoelectric parameter $g_{33}$. The constant, $ɣ_{OPG}$, takes into account the specific relationship between the tip-sample contact area and the force that the piezoelectric material can exert to the tip, and corrects the tip vibration accordingly to it. We provide a calculation of the piezoelectric stiffness and the $ɣ_{OPG}$ value for different measurement conditions. Finally, this factor is applied to the acquired data, showing a way of correcting and providing quantitative estimation of the $d_{33}$ parameter.

**Measurements into different Lead Zirconate Titanate (PZT) compositions**

Variation in Lead, Titanium and Zirconium contents of PZT composition enhances or diminishes the piezoelectric constant $d_{33}$ value[5,28]. In order to prove that quantitative measurements can be performed with PFM. We selected different PZT materials, all of them commercially available, with the following part numbers: 507 (*820*) , 505 (*610*), 503(*500*), 5A4(*460*), 5A1(*409*) and 403(*315*) provided by Morgan Advanced Materials; number in round brackets indicates the $d_{33}$ [$pm\ V^{-1}$] value, respectively[29]. Each ceramic compound is prepared for PFM measurements, following an identical polishing procedure, removing the metallic electrode (See Supplementary Figure 1 for further details). As the PZT compounds have been characterized by macroscopic measurements, a complete datasheet for each of the material is available (see Supplementary Table 2 for the full datasheet of each material). In order to calibrate the system, a Periodically Poled Lithium Niobate (PPLN) has been employed as a standard calibration pattern[30,31]. To carry out the measurements, a metallic AFM tip, with part number RMN-25PT300 and with a k constant of 18 N m$^{-1}$ has been employed[32]. Data acquisition starts by performing Amplitude vs Frequency sweeps for a range of 40 to 140 kHz, a zoom around the specific resonance of each material is depicted in **Figure 1a**.

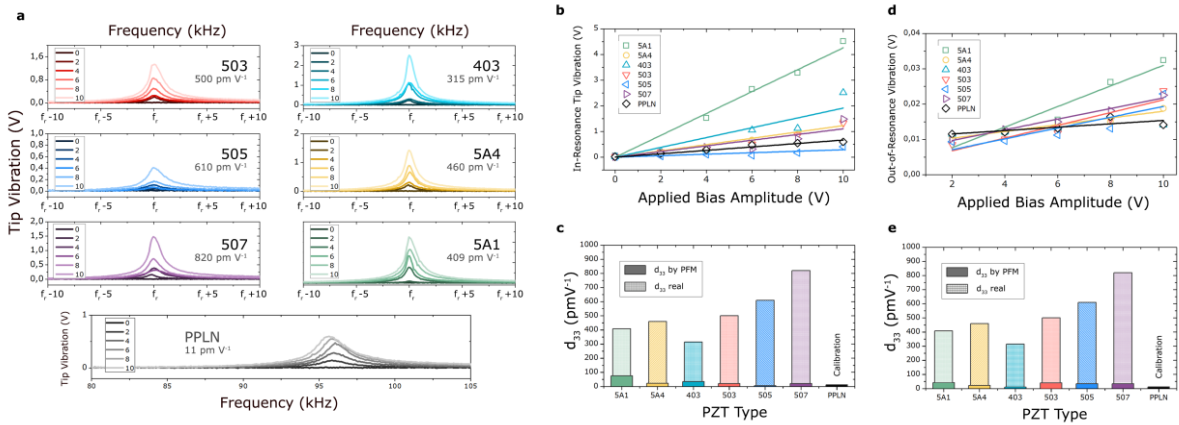

**Figure 1:** PFM measurements carried out on different Lead Zirconate Titanate materials. **a**, Tip vibration (V) vs Frequency (kHz) for 503, 403, 505, 5A4, 507, 5A1 and PPLN at different applied bias amplitude. **b**, In resonance Tip Vibration Amplitude vs Applied Bias obtained from the data in a, where a linear fit is used to obtain the $d_{33}$ value from the slope. **c**, Comparison between the real $d_{33}$ value obtained from the manufacturer datasheet and with the one measured by PFM, in resonance. **d**, Out-of-resonance Tip Vibration Amplitude vs Applied Bias for each of the material with a linear fit respectively. . **e**, Comparison between the $d_{33}$ piezoelectric constant measured by PFM and the datasheet value, out of the resonance.

Throughout the measurements, similar conditions have been selected before acquiring the data. Low humidity environment and identical measurements parameters: exact same tip used, same laser position, same force applied, same sweep rate, same Lock-In Amplifier (LIA) gain, same LIA bandwidth. From our data, we draw a full comparison of the In-resonance tip amplitude of all the materials in **Figure 1b**. As an example, the 507 PZT compound, with the highest $d_{33}$ constant, has almost the same vibration amplitude as for the case of the lowest $d_{33}$ constant material, the PPLN. A relation between the vibration amplitude and the applied bias amplitude is depicted for all the materials studied and is found linear, as expected for a piezoelectric effect[3]. By using the PPLN as a calibration material, we can assign the slope measured to a specific $d_{33}$ value of the material. By analyzing and comparing the slopes, we find the $d_{33}$ values measured with PFM, see **Figure 1c**. The data shows that the tip vibration amplitude is not directly proportional to the $d_{33}$ value as Equation (2) describes, from which we can conclude that the method is not quantitative. The measurements from Figure 1b are carried out by working at the resonance frequency. Working out-of-the resonance can be another option to see if Equation (1) is valid. In order to test this approach, we averaged the vibration amplitude for the range between 40 and 50 kHz, which is an out of the resonance measurement. The averaged value within this range, is shown in **Figure 1d**, for the case of different voltage bias amplitude. The PPLN is used as a calibration sample in order to calculate the $d_{33}$ value for the other materials.

Similar results of the in-resonance case are obtained, see **Figure 1e**, confirming that PFM is not quantitative, also for the out of the resonance working point.

**The importance of the $g_{33}$ parameter**

At this point, our data has shown that PFM measurements are not quantitative within the present standard theory. We also observed (see **Figure 1e**) that the materials having the highest $d_{33}$ constant do not vibrate as they are supposed to do. One possible explanation can be that the force exerted by the piezoelectric material is not high enough to move the cantilever. In PFM, it is assumed that the force needed to induce the tip vibration is extremely small, an assumption which, indeed, is true. Besides, the maximum force available for a piezoelectric material is directly proportional to the tip-sample contact area, which, indeed, is even smaller[33–36]. Let's take, as an example, a material with $d_{33}$ = 100 pm V$^{-1}$. The application of 10 Volts AC amplitude will induce a mechanical vibration with an amplitude of 1000 pm. Using Hooke's law, an applied force of 1,8 x 10$^{-8}$ N[37,38] can be calculated. Indeed, this force is very small, thus we may wonder if the piezoelectric material can exert such amount of force. In order to answer this question, we employ the equation available for piezoelectric actuators[33]:

$$F = \frac{A\,V}{g_{33} d} \quad (3)$$

Where F is the maximum force exerted by the material, [$N$], A is the contact area of the metallic electrode, [$m^2$], $g_{33}$ is the piezoelectric voltage constant, [$Vm\,N^{-1}$] and d is the thickness of the piezoelectric layer located between the two metallic electrodes, [$m$]. In our example, the applied voltage is 10 V while the area is estimated as a circle with a radius of 10 nm, see **Figure 2a**.

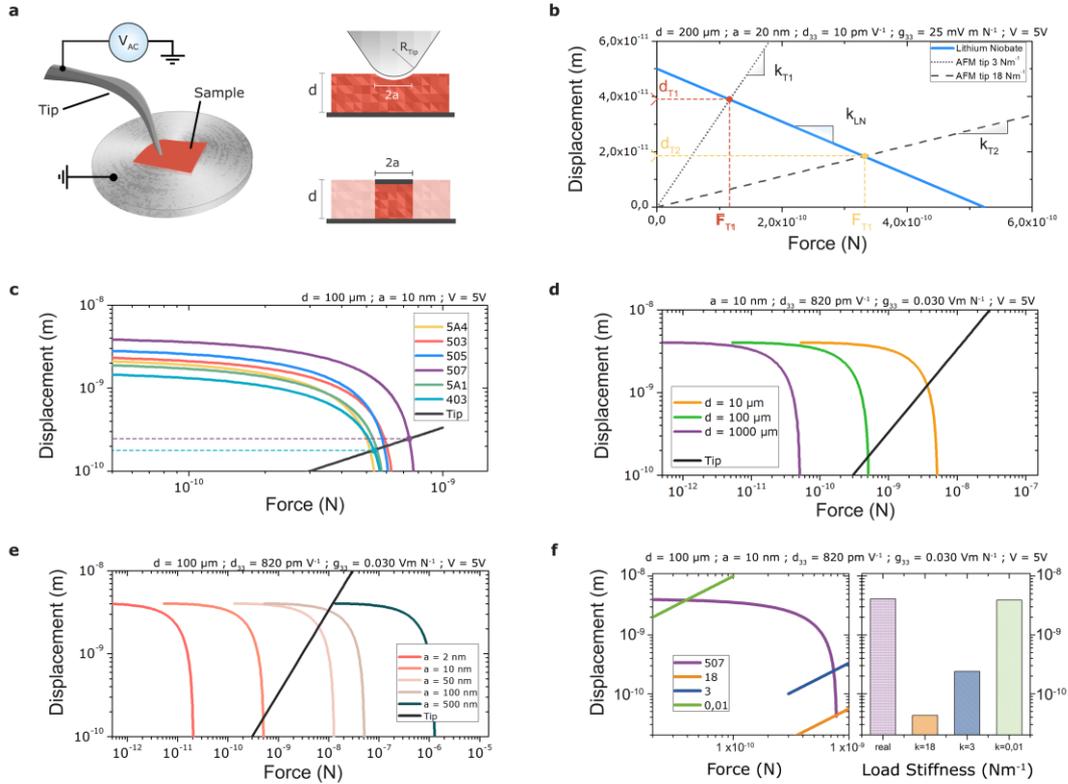

**Figure 2**: Piezoelectric stiffness comparison for a nanoscale electrode. **a** Scheme of the PFM measurement system and the tip parameters used for calculations. **b** Displacement vs Force for the case of Lithium Niobate material (blue line) where different loads are drawn with dashed and dotted line. **c** Displacement vs Force for each of the measured PZT material calculated from the macroscopic piezoelectric characteristics for a nanometric contact-lines in colors- and a load equivalent to a tip of k = 3 N m$^{-1}$ (grey line). **d** Displacement vs Force obtained for a PZT507 material, modeled for 1000, 100 and 10 µm thickness samples. **e** Displacement vs Force for different "a" parameter, creating different metallic electrode areas. **f**, Displacement vs Force curve for the case of a 507 PZT material with different loading spring stiffness (left) and measured d$_{33}$ values obtained from each of the probes (right).

The g$_{33}$ constant value of lithium niobate is an intrinsic material property with value 0.03 Vm N$^{-1}$, while the thickness, d, is 500 µm for our specific sample[31]. The maximum force the material can exert is from equation (3), 8,4 x 10$^{-10}$ N. This value represents the maximum force that the piezoelectric material can exert; hence, it occurs at a null displacement[39]. The force that the piezoelectric material can exert is much less than the one required to move the cantilever at its maximum displacement which is of 1,8 x 10$^{-8}$ N, compared with the maximum force, 8,4 x 10$^{-10}$ N. We can conclude so far that the material does not vibrate freely. If the force applied by the material is much higher, the displacement will be proportional to the d$_{33}$ value and thus this latter can be estimated correctly. On the contrary, if this condition is not respected, the relation between the force exerted by the material and the displacement plays an important role on the final underestimated value of $d_{33}$. We depicted such relation in **Figure 2b**, for the case of lithium niobate.

**Piezoelectric stiffness compared with load stiffness**

The slope of the curve in **Figure 2b** is the "piezoelectric stiffness" of the probed device ($k_{LN}$), simulated as a nanoscale size top electrode (*i.e.* the AFM probe)[40,41]. Actually, the probe on top of the material, can be modeled as a spring load[37,42]. The slope of the displacement as function of the force, is the stiffness of the cantilever ($k_T$). The intersection between the piezoelectric stiffness device and the loading stiffness is the effective working point for the system. In **Figure 2b** it can be noted that the displacement, without any force applied, should be of 50 pm, while it is reduced to 39 pm ($d_{T1}$) in presence of a 3 N m$^{-1}$ cantilever stiffness and to 18 pm ($d_{T2}$) in the case of a 18 N m$^{-1}$ one. The crossing points represent the real vibration of the tip which is measured by PFM, differing by 22% and 64%, respectively, for each probe.

We have calculated the Displacement vs Force curves for each of the materials employed in our study, the results are presented in **Figure 2c**, in a log-log scale for clarity. To obtain these curves, we used the characteristic parameters reported in the datasheet for each material, (see Supporting Table 2 of SI), where the same thickness was selected for each of the studied material. In **Figure 2c**, it is remarkable how these crossing points do not relate to the $d_{33}$ values. In fact, this crossing points for the specific tip used (k constant of 3 Nm$^{-1}$) are located in the values between 2,4 x 10$^{-10}$ and 1,7 x 10$^{-10}$ m, for all of our samples. Such values are well below the nominal values described by the standard theory in PFM, going from 4 x 10$^{-9}$ to 1,5 x 10$^{-9}$ m. It is easy to understand that the load stiffness is higher than the piezoelectric stiffness, and hence, the material will not vibrate freely. Notice, at this point, that the maximum stroke performed by a piezoelectric material depends upon other variables, among them the material thickness. In order to describe the role of the thickness into the real vibration measured by PFM, we focused on one of our sample and we explored different thicknesses (see **Figure 2d)**. In this case, the crossing points variation indicates that the thinner the sample is, the higher is the force it can exert. For all the thicknesses up to 10 μm, the force effect decreases the vibration amplitude of the piezoelectric material. In the case of thin films, both the $d_{33}$ and $g_{33}$ parameters are not the same as the datasheet values, due mainly to substrate clamping. On thin films, the $d_{33}$ values decreases, but also the dielectric constant substantially increases, diminishing the $g_{33}$ factor. Hence, in order to obtain the loading curve for ultra-thin piezoelectric film is mandatory to know both the $d_{33}$ and the $g_{33}$ piezoelectric constant values.

Another important factor to describe the electromechanical behavior in PFM measurements is the effect of the tip-sample contact area. An increase in the area results in an increased force exerted by the piezoelectric material, and hence, a better situation to work at the free vibration amplitude case (see **Figure 2e)**. However, higher area strongly decreases the lateral resolution of PFM, which is estimated in the range of several nanometers[43,44]. Increasing the force may increase the area as well, however we should also note the effect of the preloading force in the actuator, that diminishes the vibration of the piezoelectric material[44,45].

In order to reduce the cantilever stiffness effects, we calculated the working points for different loads (see **Figure 2f**) in the case of the 507 PZT sample. From the working points, we calculated the displacement value for each of the tips, obtaining the following values: 4,3 x 10$^{-11}$, 2,4 x 10$^{-10}$, 3,9 x 10$^{-9}$ m for the probe spring constants respectively of 18, 3 and 0,01 Nm$^{-1}$. Stiffer tips are currently recommended for PFM, as they help to avoid, among other, possible artifacts related to electrostatic force[46–48]. However, it strongly dimisnishes the tip vibration amplitude measured as we demonstrated before. Thus an alternative solution is mandatory to make the PFM mode quantitative also while using special engineered tips[32,49].

**Open Piezopotential Gauge**

At this point, we have demonstrated that the maximum stroke exerted by the material cannot be estimated from Equations (1) and (2). In order to include the effect in standard PFM theory, we propose the following correction for Equation (1):

$$A = d_{33} V \gamma_{OPG} \qquad (4)$$

Where $\gamma_{OPG}$ is a factor that we call "Open Piezopotential Gauge" which can be calculated as follows:

$$\gamma_{OPG} = \frac{k_a}{k_a + k_{tip}} \qquad (5)$$

Where $k_{tip}$ is the spring constant of the cantilever used and $k_a$ is the piezoelectric stiffness defined as:

$$k_a = \frac{F}{A_m} = \frac{A}{d\, d_{33}\, g_{33}} \qquad (6)$$

Where A is the tip-sample contact area (depending on the a radius), d is the piezoelectric material thickness, and $d_{33}$ and $g_{33}$ are the piezoelectric constants. This factor $k_a$ represents the piezoelectric stiffness (N m$^{-1}$) used in order to correct the amplitude measurements when out of the free vibration case. If this value is much higher that the loading stiffness, the material will vibrate freely and γ$_{OPG}$=1. In **Figure 3a**, we calculated the piezoelectric stiffness for various d$_{33}$ and g$_{33}$ parameters found in the literature.

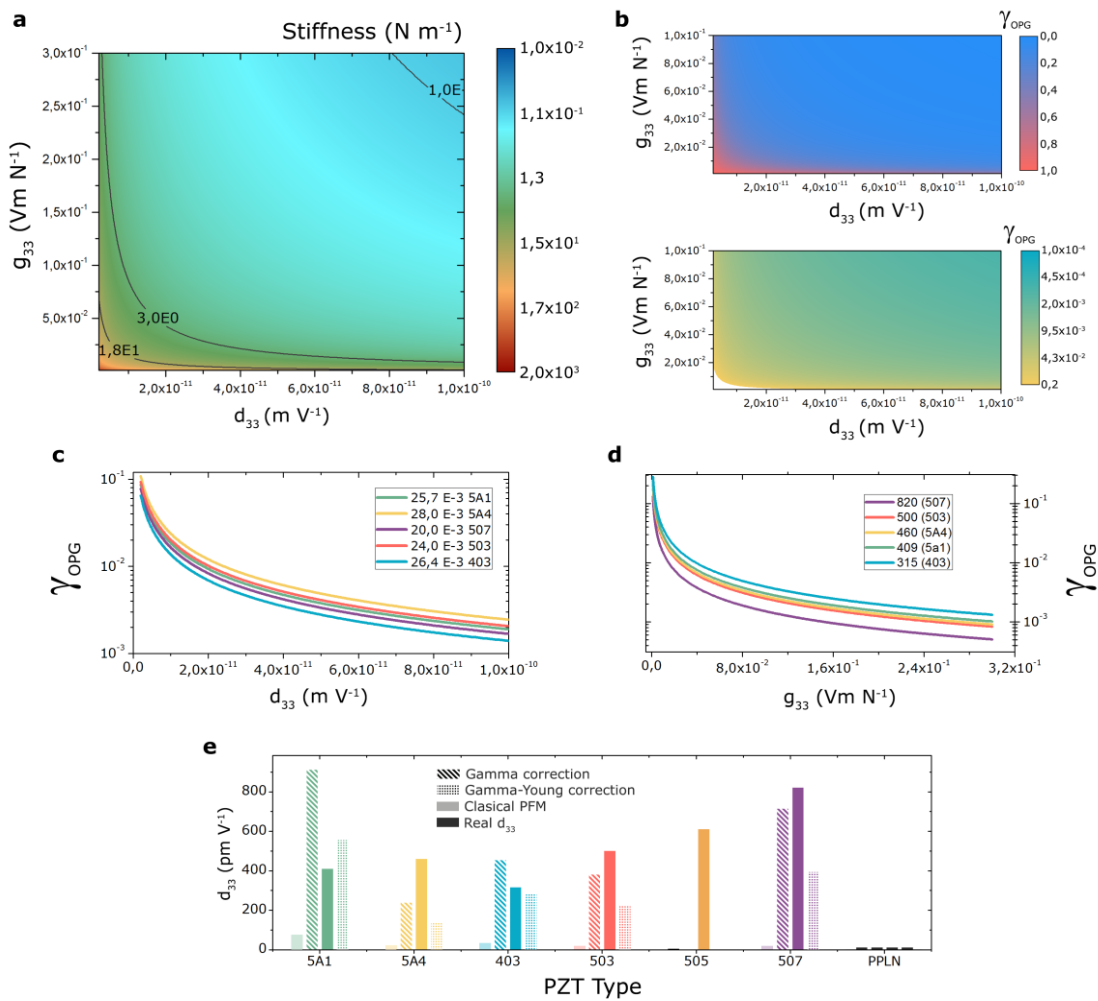

**Figure 3**: Gamma factor as a complement for classical PFM. **a.** Piezoelectric stiffness map obtained as a function of the d$_{33}$ and the g$_{33}$ piezoelectric parameters. **b**. Open Piezopotential Gauge as a function of both the g$_{33}$ and the d$_{33}$ parameters, with a contact area of radius a = 20 nm (top) and a = 5 nm (bottom). **c**. Open Piezopotential Gauge for different g$_{33}$ values-extracted from the real values of PZT materials- as a function of the d$_{33}$ parameter. **d**. Open Piezopotential Gauge for different d$_{33}$ values, as a function of g$_{33}$ parameter, for each of the PZT compounds. **e**. Corrected d$_{33}$ value obtained with the introduced gamma factor, as compared with the actual PFM measurements and the real d$_{33}$ values extracted from the manufacturer.

By performing the calculation, we find a stiffness map based upon the piezoelectric parameters. A small $d_{33}$ and $g_{33}$ value increases the stiffness of the material, while a high $d_{33}$ and high $g_{33}$ diminishes the stiffness. In the middle point we selected several cases of stiffness (*i.e.* 18, 3 and 0.01 N m$^{-1}$) as iso-load lines (black lines in **Figure 3a)**. Such lines represent the effective piezoelectric stiffness of the PFM probe spring constant. We can see that in the case of 18 N m$^{-1}$ tip, there is almost no materials that can move freely. For the case of a tip with spring constant of 3 N m$^{-1}$, it does not improve substantially. Only for the case of an ultra-soft tip, with k = 0,01 N m$^{-1}$ we can consider that the material will have the chance to vibrate freely. We thus find that the gamma factor is:

$$\gamma_{OPG} = \frac{A}{A + k_{tip} d\, d_{33}\, g_{33}} \tag{6}$$

In order to study the introduced gamma factor, we plotted the map of **Figure 3b** considering the case of a = 20 nm (top) and a = 5nm (bottom). If the gamma factor is 1, it means that the material can move freely, while if the gamma factor is close to 0, it means that the material movement is extremely dumped. In the frame of the proposed theoretical background, we see that for the majority of the cases, the gamma factor is an important damping effect into the material vibration. If we now use the Hertzian model of nanoindentation, we can find an exact value for the a parameter[44]:

$$a = \left(\frac{3P}{4E}\right)^{1/3} R_{Tip}^{1/3} \tag{7}$$

Where P is the load used, E is the Young modulus of the material and R is the tip radius of curvature. For the case of 507, E is 60 GPa, R is 20 nm for the specific tip used. Through the expression (7) we find the a parameter equals 5 nm for a force of 0,5 µN. We use now this value as a more accurate approximation to obtain the open piezopotential gauge map, which is plotted in **Figure 3b**. We can see that in the majority of the cases the gamma factor is a value much smaller than 1, meaning that the vibration amplitude of the tip is strongly damped. We can extract profiles from figure 3c, for each of the $g_{33}$ values of our PZT material (see Figure 3d). From this graph, we can see that the higher the $d_{33}$ value, the higher is the damping effect, and hence the gamma factor diminishes. If the $d_{33}$ is maintained constant, we find that for each PZT material, the $g_{33}$ plays an important role for the ɣ$_{OPG}$ factor-**see figure 3e**.

**Open piezopotential gauge as a correction factor.**
The introduced open piezopotential gauge factor can be used to correct the results obtained from PFM measurements in order to estimate the $d_{33}$ constant. For each of the PZT material, we can determine the specific $g_{33}$ value, thickness

(see Supplementary Figure 2), and, from the Young modulus and the force exerted by the tip, the tip-sample contact area. By performing the aforementioned calculation, we find the results plotted in **figure 3f**. Notably, the gamma factor corrects the values obtained from standard theory PFM and brings them closer to the real values obtained by macroscopic measurements. We specifically included two cases, where the Young Modulus is considered to calculate the *a* parameter or when the *a* parameter is maintained constant for all the measurements. If the force exerted by the piezoelectric material is considered, the values provided by PFM are much closer to the real ones. For instance, the ratio between the $d_{33}$ (measured with standard PFM) with the $d_{33}$ (from the datasheet) is 0.02 for the case of PZT507. If we now introduce the gamma factor, the ratio improves to a value of 0.9 and 0.5, for a constant *a* parameter and considering the Young modulus of the material. .

CONCLUSIONS

Piezoresponse Force Microscopy is one of the most used electrical modes available from the family of Atomic Force Microscopy techniques. It allows to study the piezoelectric effect in ferroelectric nanostructures by using a conductive nanometric tip as the local probe. In this manuscript, we performed a systematic study with different Lead Zirconate Titanate reference samples and we conclude that the PFM mode cannot be considered quantitative within state-of-the art procedure. Both resonance and out of resonance measurements depict that the tip vibration does not follow the equations of standard PFM. We specifically study the force that a piezoelectric material may exert to the tip, as a possible explanation of the quantitative controversial with this mode. The force is calculated in the frame of the standard piezoelectric theory, which is used to elucidate that the piezoelectric material should move freely. In order to accommodate the fact that this is often not the case, we introduce a correction factor, namely "Open Piezopotential Gauge", $\gamma_{OPG}$. This constant accounts for the displacement decrease of a piezoelectric material under load for conditions similar to PFM. We demonstrate that by employing this correction factor, the quantitativeness of the mode is highly improved. This method opens a new window for the possibilities of PFM as a quantitative piezoelectric characterization technique.

METHODS

**Samples preparation**
The samples are provided by Morgan Advanced Materials which are commercially available, with specific part numbers 507, 505, 503, 5A4, 5A1 and 403. Each sample consists of a ceramic piezoelectric element with two metallic contacts on each site. We polished one of the metallic contacts with different abrasive powders up to 1 µm, removing the top metallic contact. The exposed face of the ceramic element is used to perform the measurements. The bottom electrode of the sample is stuck on different steel sample holders using silver paste.

**Measurement procedure**

The same tip is used for all the measurements consisting of a Rocky Mountain Nanotechnology RMN-25PT300 tip. We specifically used the exact same probe for all the measurements, with the same laser position spot on top of the cantilever. The same LockinAmplifier (LIA) parameters are used for all the measurements: bandwidth, gain, phase offset and phase shift. Before taking data, the humidity is reduced with the use of compressed air to a value of less than 8% to avoid possible artifacts. The same preloading force for each material is used as deflection setpoint value. The measurements are obtained by acquiring Amplitude-vs-Frequency sweeps, from 40 kHz to 140kHz.

**Equipment used**

We employed a Keysight 5500 LS Atomic Force Microscopy. In roder to avoid unwanted capacitive coupling, the drive generator of the lockin is directly routed to the tip through a separate and unitary coaxial cable. The signal is routed through the break-out-box of the AFM directly to the tip.

ACKNOLEDGMENTS

We acknowledge financial support from NFFA-Europe under the EU H2020 framework programme for research and innovation under grant agreement n. 654360. ICMAB acknowledges financial support from the Spanish Ministry of Economy and Competitiveness, through the "Severo Ochoa" Programme for Centres of Excellence in R&D (SEV- 2015-0496).